\documentclass[conference]{IEEEtran}
\IEEEoverridecommandlockouts

\usepackage{float}
\usepackage{cite}
\usepackage{amsmath,amssymb,amsfonts}
\usepackage{algorithmic}
\usepackage{graphicx}
\usepackage{textcomp}
\usepackage{xcolor}
\usepackage{subcaption}
\usepackage{url}
\def\BibTeX{{\rm B\kern-.05em{\sc i\kern-.025em b}\kern-.08em
    T\kern-.1667em\lower.7ex\hbox{E}\kern-.125emX}}
\begin{document}

\title{Benchmarking Quantum Models for Time-series Forecasting\\

\thanks{Work done as part of the QCHALLenge project funded by Bundesministerium für Wirtschaft und Klimaschutz}
}

\author{\IEEEauthorblockN{Caitlin Jones}
\IEEEauthorblockA{\textit{BASF Digital Solutions GmbH} \\
 Ludwigshafen am Rhein, Germany\\
caitlin.isobel-jones@basf.com}\\
\IEEEauthorblockN{Maximilian Adler}
\IEEEauthorblockA{\textit{Aqarios GmbH} \\
Munich, Germany} 

\and
\IEEEauthorblockN{Nico Kraus}
\IEEEauthorblockA{\textit{Aqarios GmbH} \\
Munich, Germany}\\
\IEEEauthorblockN{Michael Schrödl-Baumann}
\IEEEauthorblockA{\textit{SAP SE}\\
Walldorf, Germany}
\and
\IEEEauthorblockN{ Pallavi Bhardwaj}
\IEEEauthorblockA{\textit{SAP SE}\\
Walldorf, Germany \\
} \\
\IEEEauthorblockN{David Zambrano Manrique}
\IEEEauthorblockA{\textit{Aqarios GmbH} \\
Munich, Germany}\\}
\maketitle

\begin{abstract}
Time series forecasting is a valuable tool for many applications, such as stock price predictions, demand forecasting or logistical optimization. There are many well-established statistical and machine learning models that are used for this purpose. Recently in the field of quantum machine learning many candidate models for forecasting have been proposed, however in the absence of theoretical grounds for advantage thorough benchmarking is essential for scientific evaluation. 
 To this end, we performed a benchmarking study using real data of various quantum models, both gate-based and annealing-based, comparing them to the state-of-the-art classical approaches, including extensive hyperparameter optimization.  Overall we found that the best classical models outperformed the best quantum models. Most of the quantum models were able to achieve comparable results and for one data set two quantum models outperformed the classical ARIMA model. These results serve as a useful point of comparison for the field of forecasting with quantum machine learning.
\end{abstract}


\section{Introduction}
Time-series forecasting has long been a topic of interest in the machine learning community due to its broad range of applications in finance, asset management, risk management, logistics and planning \cite{box2015time}. As this is a well established field there exist many different approaches from deep learning \cite{lim2021time}  to more heuristic methods, however as any improvement in the accuracy of a forecast may have an immediate impact, for example in the case of financial forecasting by increasing profits, the work of method development goes on. 

Quantum computing is an emerging paradigm of computing that exploits the physics of superposition and entanglement to perform calculations faster than a conventional `classical' computer\cite{48651}. There are some known quantum algorithms which offer a proven speedup over the best possible classical methods, for example \cite{shor1997,harrow2009quantum}, for certain tasks. A sub-field of quantum computing is quantum machine learning, in which quantum computers are employed as part of a machine learning workflow \cite{biamonte2017}. Currently there are no proven claims nor empirical demonstrations of robust advantage over state-of-the-art classical methods in this field, however there is a great deal of interest in its potential \cite{biamonte2017,wang2024quantum, abbas2021power}. It has been suggested that quantum computers are well suited to machine learning tasks as they are inherently probabilistic \cite{sim2019expressibility} and therefore more easily trained to model an underlying probability distribution, able to capture patterns in data using fewer parameters \cite{du2020expressive}, or better able to generalize to unseen data \cite{caro2022generalization}. In our study we focus only on the accuracy of the models, leaving aside questions of trainability or parameter counts for other work. 

Time-series forecasting has been a topic of interest in the field due to its broad applicability and the relatively low qubit count and gate depth required in many proposed methods in comparison to other machine learning problems. This is an important consideration due to the limitations of currently available quantum computers. There is a plethora of different forecasting approaches presented in the literature for both circuit architecture and annealing-based approaches. Many approaches are quantum-classical hybrid models adapted from well know classical approaches, for example: quantum neural networks \cite{IEEE2023}, recurrent quantum neural networks \cite{Siemaszko2022, bausch2020, takaki2021learning, bondarenko2023learning} and quantum reservoir computing \cite{Mujal2023, burgess2022quantum, yasuda2023quantum}.   
As forecasting is of particular interest to the financial sector, there are a large number of studies on forecasting exchange rates or share prices \cite{rivera-ruiz2022, Emmanoulopoulos2022, Thakkar2023, Cherrat2023, Srivastava2023}. Much of the literature focuses on a single quantum approach and a comparison with its classical counterpart or a state-of-the-art method. For example in \cite{Emmanoulopoulos2022}, they compare a quantum neural network to a classical BiLSTM for the task of predicting Apple stock price changes. However, as has been noted in \cite{Bowles2024BetterTC}, thorough benchmarking across multiple models is essential for a fair assessment and to avoid biasing towards positive results for quantum machine learning. 

To this end, here we investigate the potential of quantum machine learning for time series forecasting by implementing and benchmarking a selection of quantum machine learning models found in the literature. Specifically, we consider five different quantum machine learning models and compare them to three classical approaches for two real data sets.  We implemented a model optimization pipeline including k-fold cross-validation, train-evaluation-test data splitting, early stopping and hyperparameter optimization across all our models. All quantum gate models were noiselessly simulated and simulated annealing was used for the quantum annealing model.

We find that for both data sets a classical method outperformed all other methods and the relative performance of the models was highly data set dependant. Overall hyper-parameter optimization had a smaller effect than model choice, implying that, for the problems, considered exploring different model choices is more efficient than focusing on optimizing a single model. 

This paper is structured as follows: In section \ref{Background} a precise definition of the forecasting problem and a description of all of the models is given. Section \ref{Methodology} includes a description of the data sets and our experimental setup with k-fold cross-validation, early stopping and hyperparameter optimization. Section \ref{Evaluation} contains the main results of our study and a breakdown of how model performance varied with hyperparameter configuration. Finally, section \ref{Conclusion} concludes the work and offers an outlook on the field.

\section{Background} \label{Background}

\subsection{Problem definition}

A time series is defined as a set of observations \(x_t\), where \(t\) represents a specific point in time \cite{tsintroduction}. \(x_t\) can be a single value or a multidimensional vector. In this work the time series is discrete, with equidistant time steps (\(t \in \mathbb{N}\)) and the observations are univariate and real-valued (\(x_t \in \mathbb{R}\)). 

Time series forecasting aims to predict future values within the series. The predominant approach, adopted in this study, involves predicting the subsequent value \(x_{t+1}\) given the last \(n\) observations (\(x_{t-n+1}, \ldots, x_t\)). 

\subsection{ Classical Models}
 
\subsubsection{Last Value}
This baseline model uses the real value of the current time step as the prediction for the next time step.

\subsubsection{ Autoregressive Integrated Moving Average (ARIMA)}
ARIMA is a highly popular model for predicting time series \cite{box2015time}. It is autoregressive and makes the assumption of future values being dependent on past values based on a linear function. The integration deals with transforming a non-stationary time series into a stationary one by differencing. The moving average makes use of past forecast errors in the time series to predict future values.

\subsubsection{Long Short Term Memory (LSTM)}
First introduced in \cite{HochreiterLSTM} to address the vanishing gradient problem from simpler neural networks,  an LSTM model is made up of self-connected cells. Each LSTM cell has a multiplicative input gate which filters out the irrelevant input data to protect the cell state, an output gate that filters information from the current state before it is passed to other cells and a forget gate which was introduced in \cite{GersforgetgateLSTM} to filter out the obsolete information. The advantage of the LSTM over simpler architectures like recurrent neural networks is the ability to add or forget information stored in the cell state over time.

\subsection{Quantum Models}

\subsubsection{Quantum Neural Network (QNN)}
\begin{figure}[h!]
    \centering
    \includegraphics[width=\linewidth]{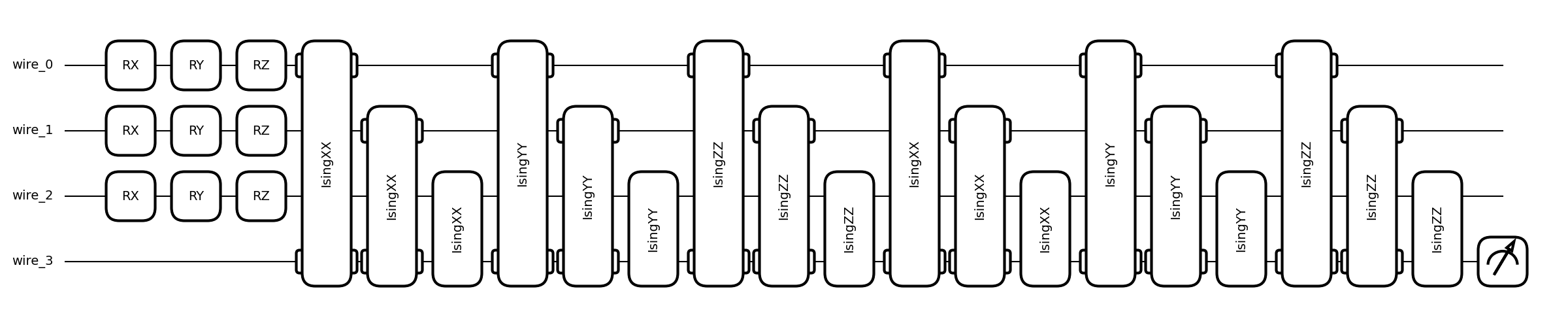}
    \caption{QNN with Ising gates}
    \label{Fig:QNN_Ising_gates}
\end{figure}

Two quantum neural networks are considered in this paper. The first QNN shown in Figure \ref{Fig:QNN_Ising_gates} is inspired from \cite{Emmanoulopoulos2022} and consists of a data encoding layer with rotation gates of Rx, Ry and Rz followed by a layer of Ising gates. Ising gates vary with circuit depth and each layer consists of IsingXX, IsingYY and IsingZZ gates.

\begin{figure}[h!]
    \centering
    \includegraphics[width=0.8\linewidth]{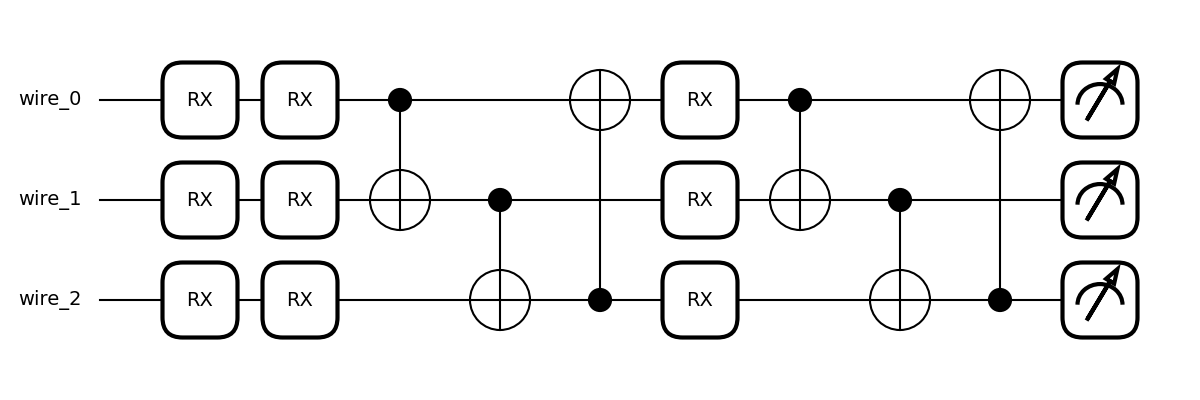}
    \caption{QNN  with pairwise entangling gates}
    \label{fig:QNN circuit}
\end{figure}

The second quantum neural network shown in Figure \ref{fig:QNN circuit} is a simple circuit where the data encoding layer consists of Rx rotation gates and the variational layers consist of Rx and pairwise CNOT gates.

\subsubsection{Quantum Deep Boltzmann Machine (QDBM)}
\begin{figure}[h!]
    \centering
    \includegraphics[width=1\linewidth]{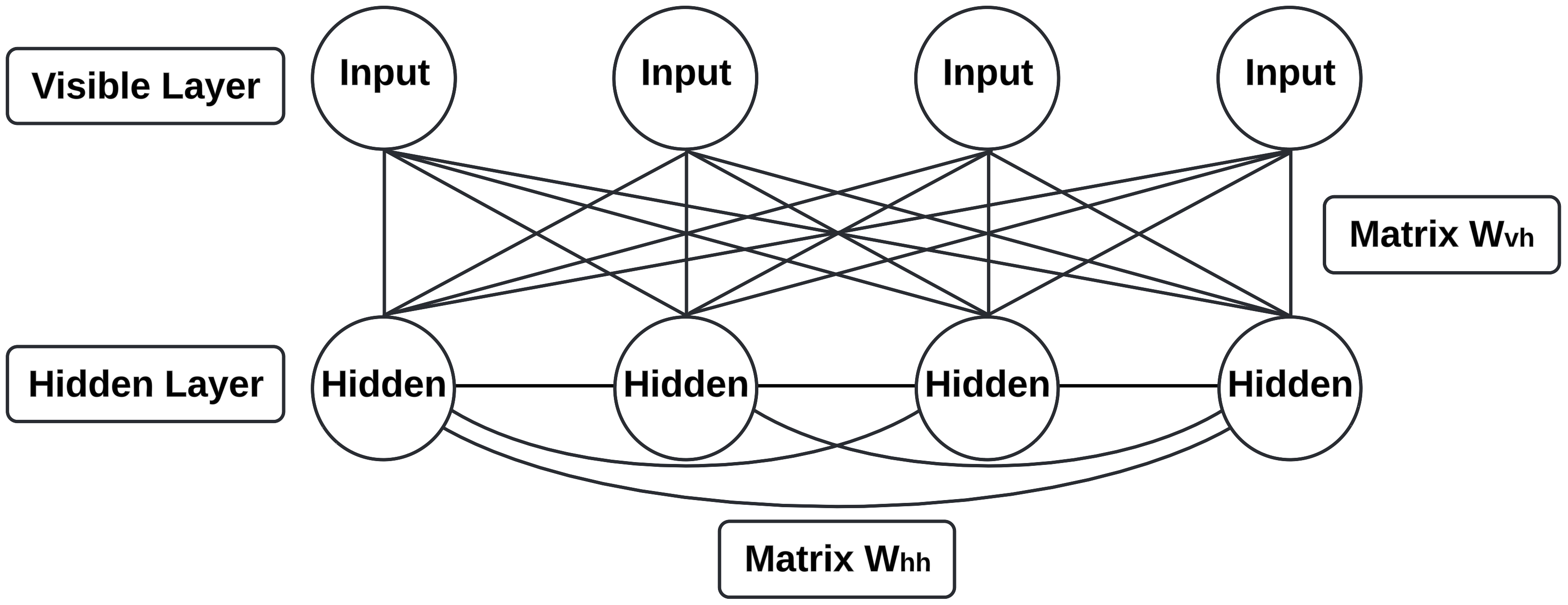}
    \caption{QDBM structure}
    \label{fig:QDBM}
\end{figure}

In this work, the Quantum Deep Boltzmann Machine originally used for reinforcement learning in \cite{qdbm} was adapted for time series forecasting. The structure (Figure \ref{fig:QDBM}) consists of a visible layer as well as a hidden layer that is fully connected to itself. It contains a visible to hidden weight matrix $W_{vh}$ and a hidden to hidden weight matrix $W_{hh}$. The nodes in the visible layer consist of the input data, while the nodes in the hidden layer are the qubits. The implementation in this work uses simulated annealing to calculate the values of its hidden layer and could potentially be done using quantum annealing. To do so a QUBO matrix is generated. The diagonal is determined by multiplying the inputs with columns of $W_{vh}$, while $W_{hh}$ is used to fill in the non-diagonal entries. The hidden to visible transformation for the output is entirely classical, thus greatly reducing the number of qubits needed. To enable the model to learn we calculate the delta of the prediction and use it alongside the learning rate and values of the hidden layer to update the two weight matrices $W_{vh}$ and $W_{hh}$.

\subsubsection{Quantum Reservoir Computing (QRC)}
Quantum Reservoir Computing uses a so-called reservoir and a readout function \cite{chen2020temporal}. In general a reservoir is given by any nonlinear input-output map with the fading memory property, for a QRC the reservoir is a fixed quantum circuit. QRC aims to exploit the exponential scaling of Hilbert space size with qubit count and the rich dynamics exhibited by dissipative quantum systems to produce an effecting mapping. The reservoir mapping is followed by a readout function to obtain the desired output. The weights of the reservoir do not require training, they are fixed parameters. The readout function is a linear function that is fitted using linear regression. 

\subsubsection{Quantum Long Short Term Memory (QLSTM)}
For the quantum-enhanced LSTM we adapted the work of \cite{chen2022quantum}, in which a Variational Quantum Circuit (VQC) replaces the classical gates responsible for information flow in a traditional LSTM. Each VQC consists of an encoding layer, then a variational layer, and lastly a measurement layer. This implementation of a quantum-enhanced LSTM is NISQ friendly, as both the number of qubits and the depth of each circuit is low. 

\section{Methodology}\label{Methodology}

\subsection{Description of the data sets}

\subsubsection{Pasta Data}
This data set was introduced in \cite{mancuso2021}. It consists of 118 daily time series of different products representing the sales of 4 pasta brands. There are also promotion flag features for each pasta brand which influences the quantity sold for that respective brand. We have aggregated all the quantities sold for all brands in one column and used it as our univariate data set for forecasting problems.

\subsubsection{Apple Stock Data}
This data set consists of the historical daily prices for Apple stock (AAPL) from Yahoo Finance. The adjusted close price was chosen for the training and prediction, it differs from the normal close price because it accounts for changes in value due to factors such as dividends, or new shares being issued.
Both data sets were min-max normalized to between 0 and 1. 

\subsection{Experimental setup}

\subsubsection{K-folds}
K-fold cross-validation is a common technique used to minimize selection bias and test a model's generalization capability for independent data sets \cite{kfolds}. Time series data has temporal dependencies, making it crucial to preserve the temporal order to prevent information leakage \cite{Cochrane_2018}. In the initial k-fold iteration, we select 500 subsequent data points, using the first 450 for training and the last 50 for comparison to the true values to find the validation error. For the next k-fold iteration, we shift the starting point by 50 time steps, incorporating 50 new unseen data points at the end for testing. Three k-folds are used for hyperparameter optimization (validation data), followed by three additional k-folds to compare each model with its best hyperparameters (testing data).

\subsubsection{Early Stopping}
Early Stopping is a method that terminates the training of a model if its validation error does not improve by a set amount within a set number of epochs. This allows models that benefit from longer training to use a higher number of epochs, without wasting time and computational resources on those models that do not. It also allows the models to adjust to each different data set during k-fold cross-validation. It can also improve the generalization of models by terminating them before convergence, thus preventing overfitting \cite{prechelt2002early}.

\subsubsection{Metrics}
The metric used for the loss function during training of the QNN models, LSTM, QLSTM and QDBM models was the mean squared error: $\text{MSE} = \frac{1}{n}\sum_{i=1}^{n} (y - \hat{y})^{2}$. The QRC model used linear regression with MSE as a measure of optimality rather than a loss function. Note that ARIMA and Last Value methods do not involve any training process. For the analysis of the results the metric used is the mean absolute error: $\text{MAE} = \frac{1}{n}\sum_{i=1}^{n} |y - \hat{y}|$. MAE was chosen as the main metric due to it being more robust to outliers and unambiguous \cite{willmott2005advantages}.

\subsection{Hyperparameter Optimization}
The goal of hyperparameter optimization was to find out which hyperparameters achieved the best performance for each data set and to have a fair comparison of the performance of each model along with the effort spent optimizing them. 
For the two data sets considered, hyperparameter optimization was carried out across each of the five quantum and two classical models using a grid search method using the training and validation data. The `Last Value' classical model was not optimized as it is a simple deterministic rule and has no parameters to be optimized. The optimization for the ARIMA model was performed with the Python module pmdarima's `Auto Arima' function \cite{autoarima}. The remaining models were manually optimized using grid search. Ideally, the number of hyperparameter configurations would be equal for each model, but the different number of relevant hyperparameters and different simulation run times of the models meant that this was not always possible, the QLSTM and QDBM models had the largest number of configurations with 108 each and the QNN and QNN Ising models had the least with 48. The hyperparameter ranges can be seen in Figure \ref{fig:overall_hyperparameters}. Each configuration was repeated ten times for each k-fold for a total of thirty runs per configuration. The best hyperparameter configuration per model per data set was taken to be the one with the lowest average MAE (averaged over all k-folds and repeated runs). These configurations were then used to train models to forecast the test data to understand the generalization performance of each model type.



\section{Evaluation} \label{Evaluation}
\begin{figure}[h]
    \centering
    \begin{subfigure}[h]{.35\textwidth}
        \centering
        \includegraphics[width=\linewidth]{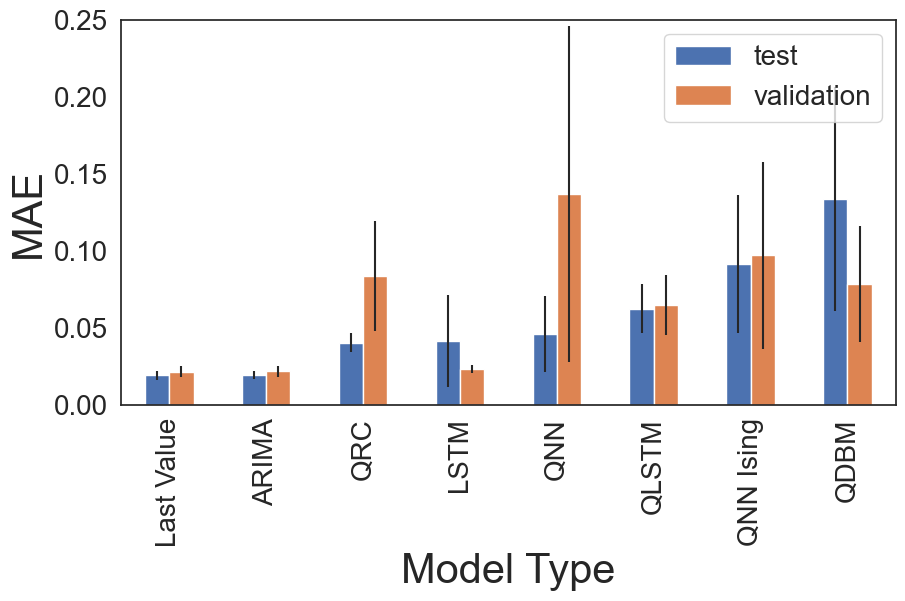}
        \caption{Apple Stock Data}
        \label{fig:Best models AAPL}
    \end{subfigure}
    
    \begin{subfigure}[h]{.35\textwidth}
        \centering
        \includegraphics[width=\linewidth]{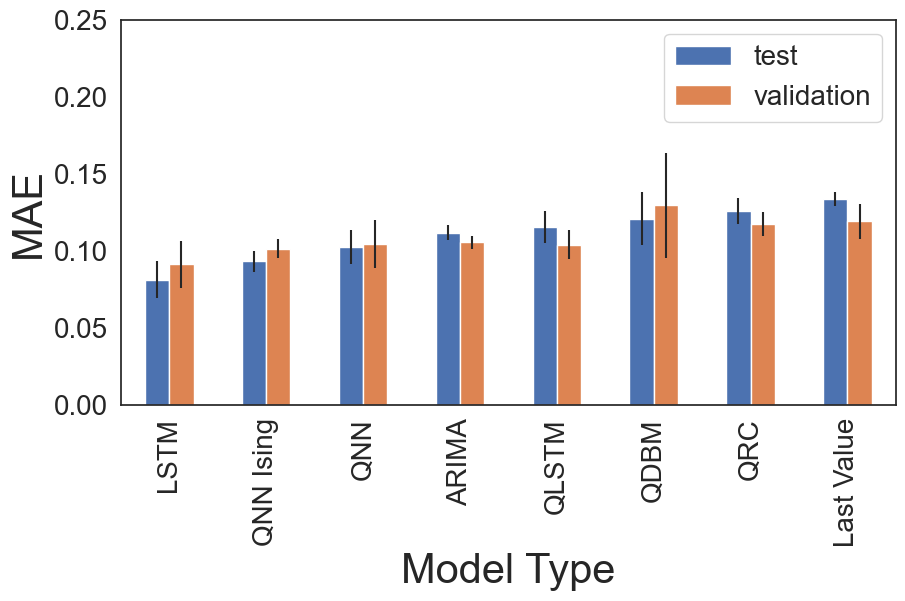}
        \caption{ Daily Pasta Sales Data Set.}
        \label{fig:Best models Pasta}
    \end{subfigure}
    \caption{Comparison of best performing models. Each bar is the mean of the performance over 10 repeated runs and 3 k-folds. The error bars are given the standard deviation of the runs.}
\label{fig:best_perf}
\end{figure}

The results for the best found model configurations are shown in Figure \ref{fig:Best models AAPL} for the Apple stock data set and in Figure \ref{fig:Best models Pasta} for the pasta sales data set. For the Apple Stock data set the simplest model `Last Value' has the lowest average error followed by the ARIMA model. The best performing quantum model is the quantum reservoir computer which had an average test error similar to that of the LSTM. It is notable that for the stock data, all machine learning models are worse than the most naive possible approach of simply taking the value at the previous time step.

For the daily pasta sales data set the situation is different, the `Last Value' model is the worst performing, and the classical LSTM model is the best performing model. The relative performance of the quantum models also varies between the two data sets, with the QRC being the best performing quantum model for the stock data but the worst performing for the pasta sales data and the inverse being true for the QNN Ising model. 

\begin{figure}[h!]
    \centering
    \includegraphics[width=\linewidth]{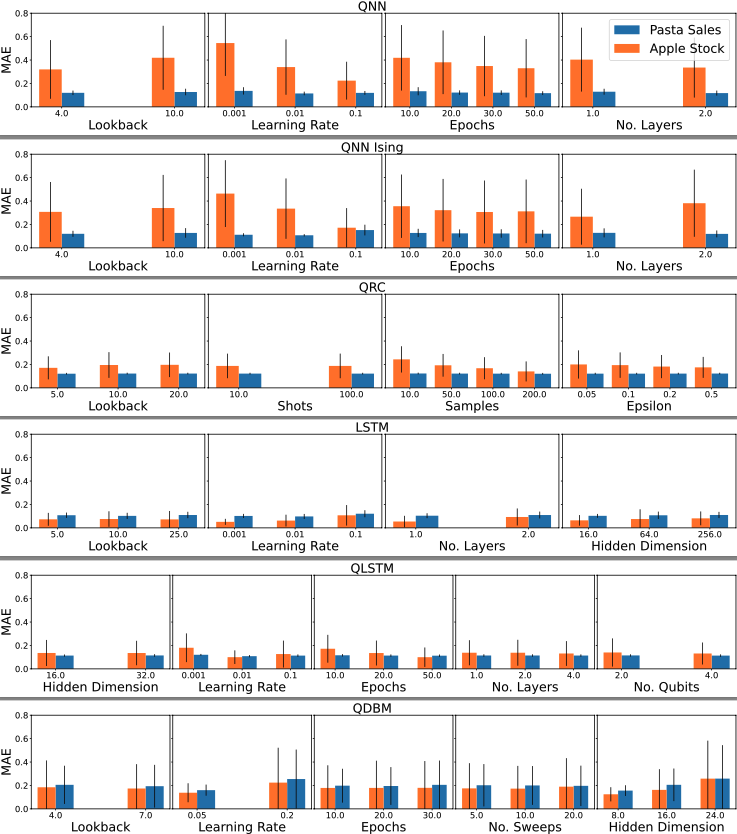}
    \caption{Comparison of MAE across hyperparameter configurations for validation data. Each row of the plot is for one model type and each subplot shows how the MAE varies for each hyperparameter (averaged over all other hyperparameters). The error bars are the standard deviation over the runs.}
    \label{fig:overall_hyperparameters}
\end{figure}

It can be seen in Figure \ref{fig:overall_hyperparameters} that the variation in performance for different hyperparameter settings is much greater for the Apple stock data set than for the pasta sales data set.  QNN and QNN Ising models had similar performance to each other for both data sets, which is likely due to their similar structure.  Although QNN Ising has more gates than QNN but their similar performance shows that more gates do not necessarily improve the performance of the model.
The only annealing-based model,  QDBM had the largest variation in performance over different hyperparameter configurations and was the only model to produce a small percentage ($<1\%$) of divergent predictions, which have been left out of the Figure.

\section{Conclusion} \label{Conclusion}
In this work, a variety of quantum machine learning models for forecasting were benchmarked against classical models for two different data sets and were compared using the mean absolute error of the forecasts. The best found configurations of the classical models had a higher average accuracy than the best quantum models. It was notable that the relative performance of the models varied greatly depending on the data set, reinforcing the fact that a quantum model offering improved performance for one data set does not imply a general advantage. The variation in performance across hyperparameter configurations also highlights the need for careful model tuning. 

Forecasting a single time step for univariate data, which we considered in this work, is one of the simplest prediction tasks.  Expanding to multidimensional input data and multiple future time steps has a broader range of applications and is left for future work, as is evaluating model performance on real quantum hardware.

\section*{Acknowledgment}

C.J. thanks Abhishek Awasthi and Davide Vodola for their assistance in setting up the supercomputing  runs for hyperparameter optimization. 

\clearpage
\bibliographystyle{unsrt}  
\bibliography{bibtex}

\begin{thebibliography}{10}

\bibitem{box2015time}
George~EP Box, Gwilym~M Jenkins, Gregory~C Reinsel, and Greta~M Ljung.
\newblock {\em Time series analysis: forecasting and control}.
\newblock John Wiley \& Sons, 2015.

\bibitem{lim2021time}
Bryan Lim and Stefan Zohren.
\newblock Time-series forecasting with deep learning: a survey.
\newblock {\em Philosophical Transactions of the Royal Society A}, 379(2194):20200209, 2021.

\bibitem{48651}
Frank Arute, Kunal Arya, Ryan Babbush, Dave Bacon, Joseph Bardin, Rami Barends, Rupak Biswas, Sergio Boixo, Fernando Brandao, David Buell, Brian Burkett, Yu~Chen, Jimmy Chen, Ben Chiaro, Roberto Collins, William Courtney, Andrew Dunsworth, Edward Farhi, Brooks Foxen, Austin Fowler, Craig~Michael Gidney, Marissa Giustina, Rob Graff, Keith Guerin, Steve Habegger, Matthew Harrigan, Michael Hartmann, Alan Ho, Markus~Rudolf Hoffmann, Trent Huang, Travis Humble, Sergei Isakov, Evan Jeffrey, Zhang Jiang, Dvir Kafri, Kostyantyn Kechedzhi, Julian Kelly, Paul Klimov, Sergey Knysh, Alexander Korotkov, Fedor Kostritsa, Dave Landhuis, Mike Lindmark, Erik Lucero, Dmitry Lyakh, Salvatore Mandrà, Jarrod~Ryan McClean, Matthew McEwen, Anthony Megrant, Xiao Mi, Kristel Michielsen, Masoud Mohseni, Josh Mutus, Ofer Naaman, Matthew Neeley, Charles Neill, Murphy~Yuezhen Niu, Eric Ostby, Andre Petukhov, John Platt, Chris Quintana, Eleanor~G. Rieffel, Pedram Roushan, Nicholas Rubin, Daniel Sank, Kevin~J. Satzinger, Vadim
  Smelyanskiy, Kevin~Jeffery Sung, Matt Trevithick, Amit Vainsencher, Benjamin Villalonga, Ted White, Z.~Jamie Yao, Ping Yeh, Adam Zalcman, Hartmut Neven, and John Martinis.
\newblock Quantum supremacy using a programmable superconducting processor.
\newblock {\em Nature}, 574:505–510, 2019.

\bibitem{shor1997}
Peter~W. Shor.
\newblock Polynomial-time algorithms for prime factorization and discrete logarithms on a quantum computer.
\newblock {\em SIAM Journal on Computing}, 26(5):1484--1509, 1997.

\bibitem{harrow2009quantum}
Aram~W Harrow, Avinatan Hassidim, and Seth Lloyd.
\newblock Quantum algorithm for linear systems of equations.
\newblock {\em Physical review letters}, 103(15):150502, 2009.

\bibitem{biamonte2017}
Jacob Biamonte, Peter Wittek, Nicola Pancotti, Patrick Rebentrost, Nathan Wiebe, and Seth Lloyd.
\newblock Quantum machine learning.
\newblock {\em Nature}, 549(7671):195--202, 2017.

\bibitem{wang2024quantum}
Yunfei Wang and Junyu Liu.
\newblock Quantum machine learning: from nisq to fault tolerance.
\newblock {\em arXiv preprint arXiv:2401.11351}, 2024.

\bibitem{abbas2021power}
Amira Abbas, David Sutter, Christa Zoufal, Aur{\'e}lien Lucchi, Alessio Figalli, and Stefan Woerner.
\newblock The power of quantum neural networks.
\newblock {\em Nature Computational Science}, 1(6):403--409, 2021.

\bibitem{sim2019expressibility}
Sukin Sim, Peter~D Johnson, and Al{\'a}n Aspuru-Guzik.
\newblock Expressibility and entangling capability of parameterized quantum circuits for hybrid quantum-classical algorithms.
\newblock {\em Advanced Quantum Technologies}, 2(12):1900070, 2019.

\bibitem{du2020expressive}
Yuxuan Du, Min-Hsiu Hsieh, Tongliang Liu, and Dacheng Tao.
\newblock Expressive power of parametrized quantum circuits.
\newblock {\em Phys. Rev. Research}, 2:033125, Jul 2020.

\bibitem{caro2022generalization}
Matthias~C Caro, Hsin-Yuan Huang, Marco Cerezo, Kunal Sharma, Andrew Sornborger, Lukasz Cincio, and Patrick~J Coles.
\newblock Generalization in quantum machine learning from few training data.
\newblock {\em Nature communications}, 13(1):4919, 2022.

\bibitem{IEEE2023}
D.~Balakrishnan, Umasree Mariappan, Pagadala Geetha~Manikanta Raghavendra, Pallela~Karthikeya Reddy, Rayavarapu Lakshmi~Narasimha Dinesh, and Shaik~Bugganapalli Jabiulla.
\newblock Quantum neural network for time series forecasting: Harnessing quantum computing's potential in predictive modeling.
\newblock In {\em 2023 2nd International Conference on Futuristic Technologies (INCOFT)}, pages 1--7, 2023.

\bibitem{Siemaszko2022}
Michał Siemaszko, Adam Buraczewski, B.~L. Saux, and Magdalena Stobi'nska.
\newblock Rapid training of quantum recurrent neural networks.
\newblock {\em Quantum Machine Intelligence}, 5:1--16, 2022.

\bibitem{bausch2020}
Johannes Bausch.
\newblock Recurrent quantum neural networks.
\newblock In H.~Larochelle, M.~Ranzato, R.~Hadsell, M.F. Balcan, and H.~Lin, editors, {\em Advances in Neural Information Processing Systems}, volume~33, pages 1368--1379. Curran Associates, Inc., 2020.

\bibitem{takaki2021learning}
Yuto Takaki, Kosuke Mitarai, Makoto Negoro, Keisuke Fujii, and Masahiro Kitagawa.
\newblock Learning temporal data with a variational quantum recurrent neural network.
\newblock {\em Physical Review A}, 103(5):052414, 2021.

\bibitem{bondarenko2023learning}
Dmytro Bondarenko, Robert Salzmann, and Viktoria-S Schmiesing.
\newblock Learning quantum processes with memory--quantum recurrent neural networks.
\newblock {\em arXiv preprint arXiv:2301.08167}, 2023.

\bibitem{Mujal2023}
Pere Mujal, Rodrigo Martínez-Peña, Gian~Luca Giorgi, Miguel~C. Soriano, and Roberta Zambrini.
\newblock Time-series quantum reservoir computing with weak and projective measurements.
\newblock {\em npj Quantum Information}, 9(16), 2023.

\bibitem{burgess2022quantum}
Adam Burgess and Marian Florescu.
\newblock Quantum reservoir computing implementations for classical and quantum problems.
\newblock {\em arXiv preprint arXiv:2211.08567}, 2022.

\bibitem{yasuda2023quantum}
Toshiki Yasuda, Yudai Suzuki, Tomoyuki Kubota, Kohei Nakajima, Qi~Gao, Wenlong Zhang, Satoshi Shimono, Hendra~I Nurdin, and Naoki Yamamoto.
\newblock Quantum reservoir computing with repeated measurements on superconducting devices.
\newblock {\em arXiv preprint arXiv:2310.06706}, 2023.

\bibitem{rivera-ruiz2022}
Mayra~Alejandra Rivera-Ruiz, Andres Mendez-Vazquez, and Jos{\'e}~Mauricio L{\'o}pez-Romero.
\newblock Time series forecasting with quantum machine learning architectures.
\newblock In Obdulia Pichardo~Lagunas, Juan Mart{\'i}nez-Miranda, and Bella Mart{\'i}nez~Seis, editors, {\em Advances in Computational Intelligence}, pages 66--82, Cham, 2022. Springer Nature Switzerland.

\bibitem{Emmanoulopoulos2022}
D.~Emmanoulopoulos and Sofija Dimoska.
\newblock Quantum machine learning in finance: Time series forecasting.
\newblock 2022.

\bibitem{Thakkar2023}
Sohum Thakkar, Skander Kazdaghli, Natansh Mathur, Iordanis Kerenidis, Andre~J. Ferreira-Martins, Samurai Brito QC~Ware Corp, Irif Universit'e~Paris Cit'e, Cnrs, and Ita'u Unibanco.
\newblock Improved financial forecasting via quantum machine learning.
\newblock {\em Quantum Mach. Intell.}, 6:27, 2023.

\bibitem{Cherrat2023}
El~Amine Cherrat, S.~Sridhar Raj, Iordanis Kerenidis, Abhishek Shekhar, Ben Wood, John Dee, Shouvanik Chakrabarti, Richard Chen, Dylan Herman, Shaohan Hu, Pierre Minssen, Ruslan Shaydulin, Yue Sun, Romina Yalovetzky, and Marco Pistoia.
\newblock Quantum deep hedging.
\newblock {\em ArXiv}, abs/2303.16585, 2023.

\bibitem{Srivastava2023}
Naman Srivastava, Gaurang Belekar, Neel Shahakar, and Aswath~Babu H.
\newblock The potential of quantum techniques for stock price prediction.
\newblock {\em 2023 IEEE International Conference on Recent Advances in Systems Science and Engineering (RASSE)}, pages 1--7, 2023.

\bibitem{Bowles2024BetterTC}
Joseph Bowles, Shahnawaz Ahmed, and Maria Schuld.
\newblock Better than classical? the subtle art of benchmarking quantum machine learning models.
\newblock {\em ArXiv}, abs/2403.07059, 2024.

\bibitem{tsintroduction}
Peter~J. Brockwell and Richard~A. Davis.
\newblock {\em Introduction to Time Series and Forecasting}.
\newblock Springer Texts in Statistics. Springer International Publishing, 2016.

\bibitem{HochreiterLSTM}
Sepp Hochreiter and Jürgen Schmidhuber.
\newblock Long short-term memory.
\newblock {\em Neural computation}, 9:1735--80, 12 1997.

\bibitem{GersforgetgateLSTM}
Felix Gers, Jürgen Schmidhuber, and Fred Cummins.
\newblock Learning to forget: Continual prediction with lstm.
\newblock {\em Neural computation}, 12:2451--71, 10 2000.

\bibitem{qdbm}
Daniel Crawford, Anna Levit, Navid Ghadermarzy, Jaspreet~S. Oberoi, and Pooya Ronagh.
\newblock Reinforcement learning using quantum boltzmann machines, 2019.

\bibitem{chen2020temporal}
Jiayin Chen, Hendra~I Nurdin, and Naoki Yamamoto.
\newblock Temporal information processing on noisy quantum computers.
\newblock {\em Physical Review Applied}, 14(2):024065, 2020.

\bibitem{chen2022quantum}
Samuel Yen-Chi Chen, Shinjae Yoo, and Yao-Lung~L Fang.
\newblock Quantum long short-term memory.
\newblock In {\em ICASSP 2022-2022 IEEE International Conference on Acoustics, Speech and Signal Processing (ICASSP)}, pages 8622--8626. IEEE, 2022.

\bibitem{mancuso2021}
Paolo Mancuso, Veronica Piccialli, and Antonio~M. Sudoso.
\newblock A machine learning approach for forecasting hierarchical time series.
\newblock {\em Expert Systems with Applications}, 182:115102, 2021.

\bibitem{kfolds}
Gavin Cawley and Nicola Talbot.
\newblock On over-fitting in model selection and subsequent selection bias in performance evaluation.
\newblock {\em Journal of Machine Learning Research}, 11:2079--2107, 07 2010.

\bibitem{Cochrane_2018}
Courtney Cochrane.
\newblock Time series nested cross-validation, May 2018.

\bibitem{prechelt2002early}
Lutz Prechelt.
\newblock Early stopping-but when?
\newblock In {\em Neural Networks: Tricks of the trade}, pages 55--69. Springer, 2002.

\bibitem{willmott2005advantages}
Cort~J Willmott and Kenji Matsuura.
\newblock Advantages of the mean absolute error (mae) over the root mean square error (rmse) in assessing average model performance.
\newblock {\em Climate research}, 30(1):79--82, 2005.

\bibitem{autoarima}
Taylor~G. Smith et~al.
\newblock {pmdarima}: Arima estimators for {Python}, 2017--.
\newblock [Online; accessed <today>].

\end{thebibliography}

\end{document}